\documentclass[11pt]{article}

\usepackage[textheight=216mm,textwidth=160mm]{geometry}
\usepackage{graphicx} 
\usepackage{natbib}
\usepackage{amsmath,amssymb,amsfonts,amsthm}
\usepackage{xcolor}
\usepackage{multirow}
\usepackage{makecell}
\usepackage{hyperref}
\hypersetup{
    colorlinks=true,
    linkcolor=blue,
    filecolor=blue,      
    urlcolor=blue,
    citecolor=blue
    }
\usepackage{anyfontsize}

\usepackage{newtxtext}
\usepackage{newtxmath}

\title{Drainage and lifetime of thin liquid films:\\ the role of salinity and convective evaporation}
\author{Tristan Aur\'egan$^1$ and Luc Deike$^{1,2}$}

\date{{\small $^1$ Department of Mechanical and Aerospace Engineering, Princeton University, Princeton, New Jersey, USA\\
$^2$High Meadows Environmental Institute, Princeton University, Princeton, New Jersey, USA}}

\begin{document}

\maketitle

\begin{abstract}

We experimentally investigate the effect of salinity and atmospheric humidity on the drainage and lifetime of thin liquid films motivated by conditions relevant to air-sea exchanges. We show that the drainage is independent of humidity and that the effect of a change in salinity is only reflected through the associated change in viscosity. On the other hand, film lifetime displays a strong dependence on humidity, with more than a tenfold increase between low and high humidities: from a few seconds to tens of minutes. Mixing the air surrounding the film also has a very important effect on lifetime, modifying its distribution and reducing the mean lifetime of the film. From estimations of the evaporation rate, we are able to derive scaling laws that describe well the evolution of lifetime with a change of humidity. Observations of the black film, close to the top where the film ruptures, reveal that this region is very sensitive to local humidity conditions.
\end{abstract}

\section{Introduction}

Thin films play a key role in ocean-atmosphere coupling: bubbles at the surface of the ocean are responsible for sea-salt aerosols generation and an important fraction of the exchanges of gas \citep{veron_ocean_2015,deike_mass_2022}. When they burst, surface bubbles eject drops of seawater in the atmospheric boundary layer, containing sea salt and any other chemical or biological contaminant that might be present in the top layer of the ocean \citep{cunliffe_sea_2013}. In particular, the sea salt aerosols generated in that way go on to influence critical climate processes such as radiative balance or cloud formation \citep{lewis_sea_2004}. Properly understanding the behavior of the bubbles at the surface of the ocean is therefore of paramount importance to model the coupling between the atmosphere and the ocean. 

Numerous studies have investigated the behavior of surface bubbles, with the goal of finding a relationship between the bubble size and the size distribution of the ejected drops \citep{lhuissier_bursting_2012,anguelova_effects_2017,poulain_ageing_2018,berny_statistics_2021,shaw_film_2024,jiang_submicron_2022,jiang_abyss_2024,deike_mechanistic_2022}. Several pathways for drop formation have been identified depending on bubble size with associated distributions, however several questions remain for the complete modeling of the drop generation over the ocean: first, the role of salt on the drainage or lifetime of surface bubbles is still uncertain. It is not accounted for in the models but has a visible effect on bubble bursting as demonstrated by several lab experiments \citep{martensson_laboratory_2003,park_mixing_2014,may_lake_2016,zinke_effect_2022,dubitsky_effects_2023,mazzatenta_linking_2024}. 

An additional difficulty in the study of surface bubbles is the link between drainage and lifetime, i.e. identifying a single rupture criterion. For many of the drops ejected, the typical size is given by the thickness of the cap at the instant of burst \citep{lhuissier_bursting_2012,poulain_ageing_2018,shaw_film_2024} such that both information (thickness over time and time of burst) are important. However, while drainage is deterministic and can be modeled as a function of a few parameters, the lifetime of surface bubbles is intrinsically stochastic and exhibits wide distributions. \citet{poulain_ageing_2018} discussed in their introduction the difficulty of obtaining these distributions even in laboratory settings as seemingly without changing experimental parameters, the lifetime can change drastically. Finding a link between drainage and lifetime fluctuations would therefore improve our understanding of those bubbles.

Finally, the experiments described above have all been performed in a still (closed box) or uncontrolled atmosphere. Yet, the air surrounding the bubble plays a role in the drainage through the humidity \citep{poulain_ageing_2018,pasquet_impact_2022} and in the bursting process through the air density \citep{jiang_submicron_2022}. In realistic oceanic conditions, the bubbles move at the surface and are subjected to the wind: airflow over the cap of the bubbles may change the drainage laws or the lifetime distribution through several mechanisms. First, it may entrain the film \citep{burgess_instability_1999}, changing the flow structure inside of the bubble cap and therefore altering drainage. Second, airflow can change the evaporative rate, by forcing the convection and replacing the natural convection that takes place around surface bubbles in quiescent environments \citep{dietrich_role_2016,boulogne_convective_2018}.

We set out to investigate the effect of water salinity, atmospheric humidity and airflow on thin films to identify how these parameters affect the drainage and burst. Here we start by considering flat soap films instead of the cap of surface bubbles as a model object to investigate the effect of humidity and salinity. Soap films present the advantage of a better controlled generation and geometry allowing for a very repeatable initial condition. They also present the benefit of a direct visualization of the motion in the film at all times through interferometry. 
The films however require large surfactant concentrations to be stable for a few seconds. As a consequence of this large surfactant concentration, soap films usually rupture near the top \citep{saulnier_study_2014,pasquet_lifetime_2024}, in contrast to low contamination bubbles, which can burst anywhere in the cap \citep{champougny_life_2016}.

Soap film drainage and lifetime has been the focus of many theoretical \citep{naire_insoluble_2000,degennes_young_2001,champougny_surfactantinduced_2015} and experimental studies \citep{mysels_soap_1964,langevin_thinning_1994,berg_experimental_2005,sett_gravitational_2013,seiwert_velocity_2017,monier_selfsimilar_2024}. Despite these efforts, there is no general law describing the drainage of the film, depending on the fluid properties and type of surfactant used. Several fitting laws have however been used to describe the evolution of the thickness \citep{berg_experimental_2005, monier_selfsimilar_2024}. Film drainage and lifetime are closely linked, but in a non-trivial way: \citet{champougny_influence_2018} studied the effect of both in the context of continuously pulled soap films and found that while the thickness of the film is independent of humidity, the lifetime strongly depends on it. They also propose a mechanism of film evaporation that rationalizes this observation. \citet{pasquet_lifetime_2024} also studied the effect of evaporation on lifetime in giant soap films with interfaces rigidified by surfactants. 

In this paper, we first describe qualitatively the effect of salinity on thin films. We then investigate the drainage, varying salinity and humidity. We show that humidity does not influence film drainage and that the effect of salinity only appears through its effect on fluid viscosity. We finally focus on the influence of humidity on lifetime, showing the importance of mixing in the atmosphere around the film and exhibit scaling laws describing the evolution of lifetime with humidity. 

\section{Experimental setup}\label{sec:setup}

\begin{figure}
    \centering
    \includegraphics[width = \textwidth]{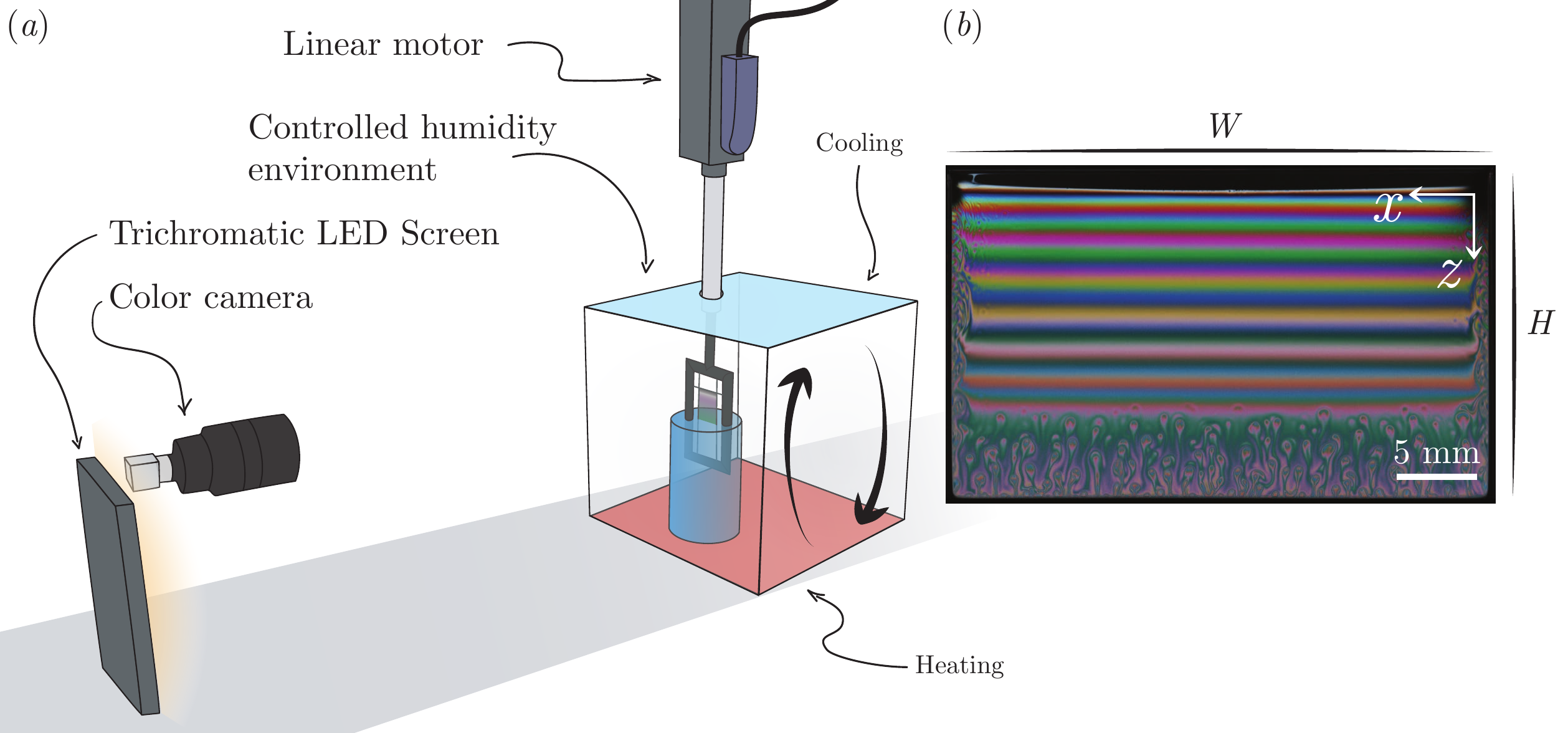}
    \caption{(a) Setup used to generate soap films in an environment with controlled humidity. (b) Snapshot of the interference pattern obtained. This film is generated in the case without salt and with 70\% relative humidity.}
    \label{fig:setup}
\end{figure}

The experiments are designed to generate soap films from a solution bath and to allow them to drain until they spontaneously burst. The films are held up by a frame made out of three wires (Nylon, diameter 100 \textmu m) and their bottom border is connected to the bath (the detailed setup is shown in Fig. \ref{fig:setup} (a)). The wires are attached to a 3D printed rigid frame and the films have a width of 29 mm and a height of 20 mm. The rigid frame is attached to a linear motor, allowing for a precisely controlled and repeatable generation. The films are pulled out of the solution with a constant velocity of 1 cm/s and brought to a stop gradually (using a constant deceleration of -2.5 cm/s$^2$) at the desired height. After generation, the films are held in place until they spontaneously burst: this event is automatically detected by the camera and the linear motor the lowers the frame back in the bath. After a short delay of 5~s, a new experiment can start.

The bath and film are placed in a chamber where the relative air humidity is precisely controlled. The setup used to achieve this control is very similar to the one described in great detail in \citet{boulogne_cheap_2019} and is able to maintain the humidity within $\pm 1\%$. A three-way valve alternatively injects humid or dry air in the chamber. This valve is duty-cycle controlled using a humidity measurement inside the chamber (sensor HIH-4021-003). 

The thickness is recorded via color thin film interferometry \citep{atkins_investigating_2010,sett_gravitational_2013}. The film is illuminated with a quasi-normal incident trichromatic light (Fig. \ref{fig:setup} (b)). This incident light is generated by three LED strips with narrow bandwidths (typically 15~nm) and a main wavelength in the red, green and blue, respectively (SimpleColor\texttrademark LED Strips by Waveform Lighting). A color camera then records the interference pattern, and by separating the three color channels, three quasi-monochromatic interference patterns are recovered. Using a color camera allows us to locate the first (white) fringe and then count the fringes on any of the three monochromatic patterns to recover the thickness anywhere in space or time. Images of the film draining are recorded at 25 frames per second and the color interference technique allows us to track the thickness starting from a few seconds after the film has stopped being pulled out.

The solutions used to generate soap films were prepared with deionized water, adding NaCl in amounts ranging from 1 g/L to 147 g/L. A non-ionic surfactant (Triton X-100) was added with a concentration of three times the Critical Micelle Concentration (CMC) to stabilize the film. This surfactant was chosen to minimize the interactions between the dissolved ions and the surfactant. The static surface tension of the various solutions was measured using a Langmuir trough and was found to be essentially independent of the salinity and equal to 32~mN/m (see App. \ref{app:langmuir}). At high concentrations, solutions of NaCl have a viscosity up to 2 times that of pure water \citep{kestin_tables_1981}, to isolate the effect of viscosity, we therefore also tested solutions of 5 and 10\% Glycerol by weight with the same surfactant. A summary of the experimental conditions is provided in Tab \ref{tab:conditions} and is separated in two parts: in the drainage part, the experiments are repeated five to ten times, and we explore many different conditions. Note that we also performed experiments with a different concentration of Triton X-100 (1 CMC) and with C12E9, another non-ionic surfactant. In the lifetime section, we focus on obtaining robust statistics in the zero salinity case, but changing the relative humidity.

\begin{table}
    \centering
    \begin{tabular}{c c c c c c c}
        & Surfactant & \makecell{Salinity (g/L)} & \makecell{Glycerol (\%wt)} & \makecell{Relative \\humidity (\%)} & \makecell{Convection ($^\circ C$)} & Repetitions \\ 
       \multirow{4}{*}[-2em]{\rotatebox[origin=c]{90}{Drainage}} & \makecell{Triton X-100 \\(3 CMC)} & \makecell{0, 1, 3.5, \\11, 35, 72,\\ 147} & 0 & \makecell{30, 50, \\ 70, 90} & 0 & 5 - 10 \\ 
        & // & 0 & 5, 10 & \makecell{30, 50, \\ 70, 90} & 0 & 5 - 10 \\ 
        & \makecell{Triton X-100 \\(1 CMC)} & 35 & 0 & \makecell{30, 50, \\ 70, 90} & 0 & 5 - 10 \\ 
        & \makecell{C12E9 \\(3 CMC)} & 0, 147 & 0 & \makecell{30, 50, 70} & 0 & 5 - 10 \\ \hline 
       \multirow{2}{*}{\rotatebox[origin=c]{90}{Lifetime}} & \makecell{Triton X-100 \\(3 CMC)} & 0 & 0 & \makecell{30, 40, 50, 60, \\ 70, 80, 90} & 0, 5 & 100 - 200 \\ 
        & // & 0 & 0 & 80 & \makecell{0, 3, 5, \\ 7, 10, 15} & 100
       \end{tabular}
    \caption{Summary of experimental conditions. Experiments with surfactants C12E9 and Triton X-100 at 1 CMC have been done for comparison but are not shown on the graphs for clarity.}
    \label{tab:conditions}
\end{table}

In order to explore the effect of mixing of the air in the chamber, we use Rayleigh-Bénard convection. We use a heating mat at the bottom of the chamber and Peltier plates at the top to achieve the desired temperature difference. The temperatures at the top and bottom surface are measured and controlled using analog sensors (Texas Instruments LM35DZ/LFT4), and the temperature at the level of the soap film is also measured to ensure that it remains between 21 and 22$^\circ$C. The heating mat and Peltier plates are duty-cycle controlled such that the wall temperature is constant. The Rayleigh number controlling the instability is $Ra = \rho_a \beta_a \Delta T \ell^3 g / \mu_a \alpha_a = \mathcal{O}(10^7)$, with $\ell = 60$~cm the height of the chamber, $\beta$ the thermal expansion coefficient, $\rho_a$ the density of air, $\mu_a$ the viscosity of air, $\alpha_a$ the thermal diffusivity of air, and a temperature difference $\Delta T$. At this Rayleigh number, the flow in the chamber is turbulent \citep{drazin_rayleighbenard_2002} and the air surrounding the film is therefore homogeneous in space as any slow or local variations of the experimental conditions are suppressed. The instability sets in air in the chamber into motion, and the typical root-mean-square velocity is 5~mm/s (measured using Particle Image Velocimetry). In the case where mixing is not activated, the air in the chamber can still have some motion, but it is much slower: the air inlets are placed at the bottom of the chamber and a diffuser restricts the air inlet flow velocity. A small ($< 5$\%) but stable vertical humidity gradient forms in the chamber, these effects resulting in a velocity too small to measure with our PIV setup ($< 1$~mm/s).
We detail the effect of mixing the air in the chamber in Sec. \ref{sec:lifetime}; in the drainage section (Sec. \ref{sec:drainage}) there is no mixing, but we show in Fig. \ref{fig:fluc_lifetime} that convection does not alter the drainage significantly.

\section{Qualitative observations}\label{sec:quali}

\begin{figure}
    \centering
    \includegraphics[width = \linewidth]{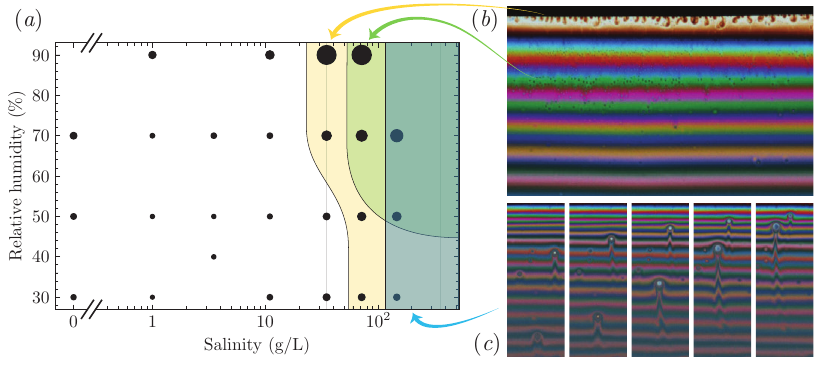}
    \caption{(a) Map of the experimental data points in the relative humidity and salinity space. The size of the dots is proportional to mean lifetime across repetitions. The gray vertical line at 35 g/L is the salinity of sea water, the one at 360 g/L is the solubility of NaCl. The three colored areas are the regions where each of the features is visible: rising patches in the film in blue, small falling particles in green and Rayleigh-Taylor like instability at the top of the film in yellow. \quad (b) Illustration of the top instability (brown patches falling in the white fringe) and of the small patches falling down in the center of the film. \quad (c) Illustration of the rising patches at high salinity. Each image is separated by 0.2~s. In both of these illustrations, salinity is 147 g/L and relative humidity is 50~\%. }
    \label{fig:quali}
\end{figure}

Using our video recordings of the films draining with variations over two orders of magnitude of salinity, we compare qualitatively the film drainage to its equivalent without salt. The features visible in the draining films are unchanged up to salinities close to seawater concentrations. For these lower concentrations, the images of the film look identical to the case without salt, presented in Fig. \ref{fig:setup} (b). At higher concentrations, qualitatively different features start to appear. They are summed up in Fig. \ref{fig:quali} and placed in the parameter space of relative humidity / salinity. The colored area corresponds to parameters where a given feature occurs at least 50\% of the repetitions of the experiment. We briefly describe below those features and the cases where they appear. 

From moderate salinities and at any humidity, an instability is occurring near the top of the film. There, thick patches are formed at the interface between the colored film and the black film (brown plumes at the top of Fig. \ref{fig:quali} (b) and yellow area in panel (a)). These patches are thicker than the ones surrounding them and therefore fall down in the film because of buoyancy until they reach the level that matches their thickness. This instability is reminiscent of a Rayleigh-Taylor instability, or to the one observed by \citet{shabalina_rayleightaylorlike_2019}. This instability occurs in many of the cases studied: with other surfactants and in cases where glycerol was used instead of salt. It is therefore not specific to salted films and most likely simply a feature of long-lasting films and was previously studied in different contexts \citep{seimiya_revisiting_2021}.

From salinities about twice that of seawater, very small patches (smaller than 100~µm) can be seen throughout the upper region of the films (dots in Fig. \ref{fig:quali} (b) and green area in panel (a)). These patches seem to be generated in the bulk of the film and are heavier than their surroundings as they can be seen falling down in the film. In contrast to the previous observation, this feature can't be seen with glycerol alone. We also could not observe the same phenomenon in the few cases with the surfactant C12E9 that we performed. These patches may be salt crystals generated near the top because of the evaporation occurring preferentially near the top of the film and therefore locally increasing the salt concentration past its solubility. Similar features are believed to have been spotted in surface bubbles \citep{poulain_ageing_2018}. The fact that we couldn't observe it with another surfactant may also point towards a particle that is formed only in the presence of salt and Triton X-100: some ionic surfactants are for instance known to form solid phases in the presence of NaCl \citep{kharlamova_interfacetemplated_2024}.

For extremely high salinities (four times seawater), thin patches can be seen rising from the middle of the film upwards (Fig. \ref{fig:quali} (c) and blue area in panel (a)). These patches are generated preferentially at the beginning of the film lifetime when it is still thick. They are not visible with glycerol alone, and can also be found with other surfactants. By exchanging fluid between the middle of the film and its top, these patches may alter the drainage dynamics of the film.

\section{Drainage}\label{sec:drainage}

\begin{figure}
    \centering
    \includegraphics[width = \textwidth]{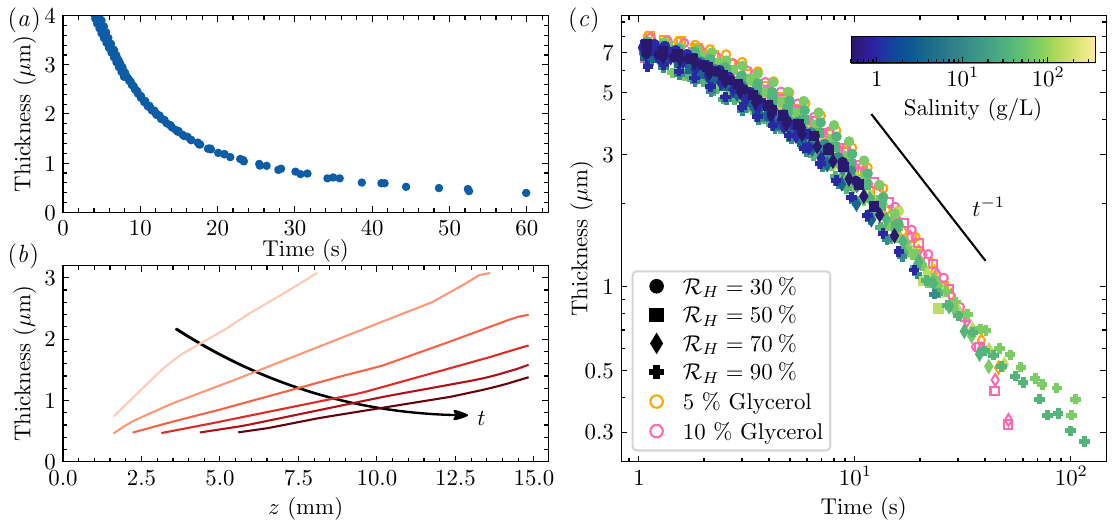}
    \caption{(a) Typical evolution of the thickness over time at a fixed vertical location in the film (5 mm below the top wire). \quad (b) Vertical profiles of thickness at several instants throughout the lifetime of the film. The data shown in (a) and (b) corresponds to a film with a salinity of 72 g/L and a relative humidity of 70 \%. \quad (c) Thickness over time for all solutions and humidities measured. Markers show the relative humidity and the colorbar shows the salinity of the solution. In addition, two experiments with Glycerol but no NaCl are also displayed (orange and pink markers with 5 and 10 \% Glycerol by weight, respectively). Each marker corresponds to an average between 5 to 10 repetitions.}
    \label{fig:drainage}
\end{figure}

Using the videos of the film draining and the interferometry technique described in Sec. \ref{sec:setup} we are able to measure the thickness of any given point, over time. A typical drainage curve for all the cases we explore in shown in Fig. \ref{fig:drainage} (a): the thickness is a few microns and drains over a timescale ranging from a few tens of seconds to minutes. We can also recover vertical thickness profiles by interpolating between consecutive interference fringes, at a given instant. This technique works from a few seconds after the film has stopped being pulled out ($t \gtrsim 2$ s) and outside the regions where marginal regeneration patches are present ($z \lesssim 15$ mm). This allows us to investigate how the vertical profile of thickness evolves over time (illustrated in Fig. \ref{fig:drainage} (b)). The thickness increases almost linearly when moving towards the bottom of the film, with a slope $\partial h / \partial z$ that decreases in time.


Within the parameter range tested, film drainage is remarkably similar: without any rescaling the drainage curves almost collapse onto a single curve (Fig. \ref{fig:drainage} (c)). In particular the fact that the drainage curves for different humidities are the same (different markers) shows that in the regime of interest with films lasting at most a few hundred seconds, evaporation has a negligible role in the thinning. This property (which was previously noted in the case of continuously drawn films \cite{champougny_influence_2018}) holds true regardless of salinity, despite the known effect of salt (or glycerol) to alter evaporation rates \citep{el-dessouky_evaporation_2002,roux_everlasting_2022a}. 
The effect of salinity (different colors), is to slightly shift the curves without altering the initial thickness of the film. The rate at which the film drains is therefore altered by salinity, but this change remains moderate in the conditions tested. Similarly, increasing the viscosity (orange and pink markers) slightly shifts the curves and slows down drainage. At very long times, the curves corresponding to Glycerol and NaCl solutions start to deviate from each other. We interpret this result as being due to the qualitatively different features visible in the films.
The collapse of the drainage is robust to a change of surfactant and / or surfactant concentration in the cases we explored (not shown on Fig. \ref{fig:drainage}). 

At long times (after about 10~s), the drainage curve exhibits a power law behavior in time with an exponent close to -1 (i.e. $h \propto 1 / t$). \citet{monier_selfsimilar_2024} recently proposed a general form for the evolution of thickness at a given height $z^*$ in the film: $h(z^{*}, t) \propto (\kappa / (t - t_0))^m$ with $\kappa$, $t_0$ and $m$ constants that can vary slightly depending on the experiment. Using data from the literature as well as a series of experiment with various viscosities and frame geometries, they found an exponent $m$ between 1 and 2, with an average value of 1.41. By fitting the same law onto our present experiment, we find an exponent $m = 1.3 \pm 0.2$. As discussed in their work, we also find a small negative value for $t_0$ corresponding to the initial finite thickness of the films. The exponent found here is coherent with previous experiments on soap films where evaporation does not play a major role and where the main draining mechanism is the exchange of patches of fluid with the Plateau borders at the sides of the film \citep{berg_experimental_2005}. These films are said to be mobile \citep{mysels_soap_1964} in opposition to rigid films where the interface is contaminated to the point where it applies a no-slip boundary condition to the interstitial fluid. In that case, the equation describing the thinning of the film is \citep{mysels_soap_1959}: $
    \partial h/ \partial t = - \rho g/ 4\mu \times h^2 \partial h/ \partial z $,
with $\mu$ and $\rho$ the viscosity and density of the liquid, respectively. A downwards Poiseuille flow develops inside of the film, leading to a thickness varying like $h \sim \sqrt{z/t}$, clearly not compatible with the present experiments. The surfactant gradient stabilizing the film can instead be taken into account by introducing a finite stress at the interface. This stress \emph{a priori} depends on the surfactant used and on the surfactant concentration and should be computed by introducing the equation of state for the surface excess of surfactant \citep{champougny_surfactantinduced_2015}. Under the simplifying assumption of a constant slip length (or stress) at the interface, previous authors have numerically recovered a thinning law close to $t^{-1}$ \citep{berg_experimental_2005} with an exponent that depends on the choice of boundary condition. The variations of the drainage exponent $m$ visible in the experiment can therefore be interpreted as a change of boundary condition for different cases.


\begin{figure}
    \centering
    \includegraphics[width = \textwidth]{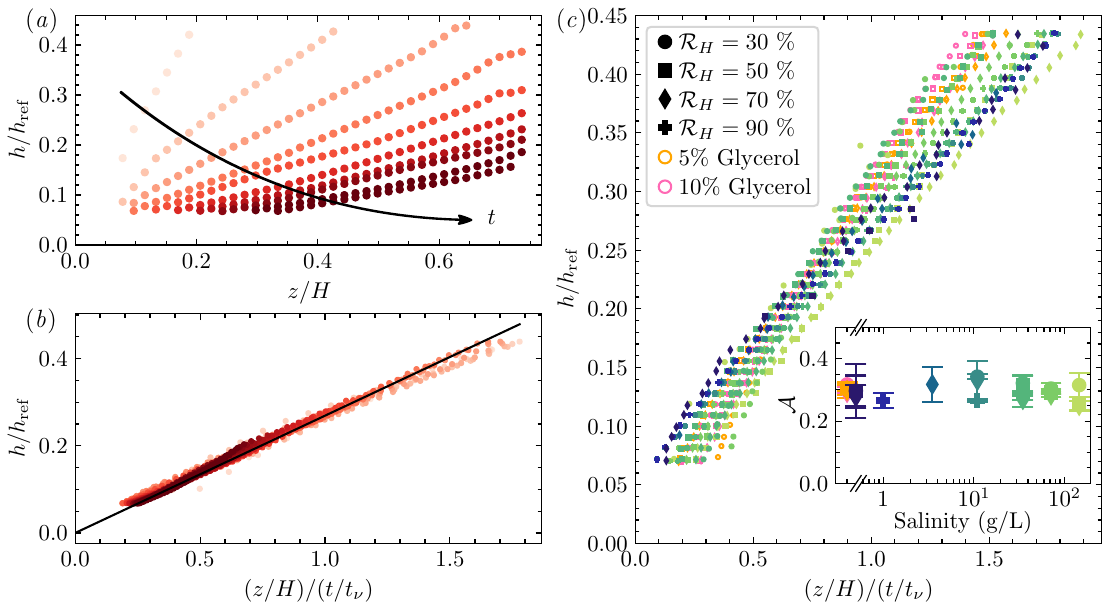}
    \caption{(a) Dimensionless thickness as a function of $z / H$, color indicates time. \quad (b) Thickness as a function of the rescaled parameter following Eq. \ref{eq:hzt} The data shown in (a) and (b) corresponds to a film with a salinity of 72 g/L and a relative humidity of 70 \%. The black line is a linear fit with slope 2.0. \quad (c) Thickness as a function of the rescaled parameter, for all salinities and relative humidities tested. Colors and markers are identical to Fig. \ref{fig:drainage}: markers indicate relative humidity and colors indicate salinity or Glycerol concentration. Each marker corresponds to an average between 5 and 10 repetitions. (Inset) Measured proportionality constant as a function of the salinity (Eq. \eqref{eq:hzt}). Errorbars represent the dispersion between repetitions.}
    \label{fig:hzt}
\end{figure}

When plotted as a function of $z / t$, the thickness of the film follows a linear trend (i.e. $h(z, t) \propto z / t$, see Fig. \ref{fig:hzt} (b)). From a few seconds after the film generation and for vertical positions above the one influenced by marginal regeneration, the thickness obeys a linear relationship with $z/t$ with no fitting parameters. This strikingly simple result is robust to changes of surfactant and salinity (or viscosity) within our parameter range (Fig. \ref{fig:hzt} (c)). 

The slope of this linear relationship $\mathcal{A}$ is defined using
\begin{equation}\label{eq:hzt}
    h(z, t) / h_{\rm ref} = \mathcal{A} \frac{t_{\nu}}{H}\frac{z}{t} ,
\end{equation}
with $h_{\rm ref} = 7$ \textmu m, the thickness at generation and $t_{\nu}$ the viscous drainage time. This typical timescale can be obtained by considering the thin film equation \citep{mysels_soap_1959}, giving:
\begin{equation}
    t_{\nu} = \frac{\mu H}{\rho g h_{\rm ref}^2}.
\end{equation}  
For the solutions considered here this typical viscous time is 40 to 50~s depending on the salinity. Using the drainage scaling in combination with the viscous timescale, we can see that the thickness data for all solutions reasonably collapses onto a straight line (Fig. \ref{fig:hzt} (c)). The slope $\mathcal{A}$ of this linear relationship is an order $\mathcal{O}(1)$ non-dimensional constant, which can be evaluated and is independent on the salinity or on the viscosity ($\mathcal{A}\approx 0.3$, inset). Using glycerol or NaCl, the effect of the viscosity is therefore fully accounted for through the viscous timescale $t_\nu$. 

There is a small visible offset between different solutions that may be attributed to the drainage at a fixed height following a power law close to but not exactly minus one or the shape vertical profile not being always perfectly linear. Using the fitted exponent $m$ and the more general expression proposed by \citet{monier_selfsimilar_2024} to rescale the data yields a very similar result to the one plotted on Fig \ref{fig:hzt} (c), but our approach presents the advantage of allowing us to measure the slope $\mathcal{A}$ without any fitting parameters, and this single parameter is an order 1 constant.

The drainage scaling presented here is robust for all conditions tested of salinity and relative humidity but also to a change of surfactant concentration or surfactant (Triton X-100 at 1 CMC or C12E9 at 3 CMC). This law is only valid from a few seconds after the start of the drainage (Fig. \ref{fig:drainage} (c)) and applies only in the bulk of the film, away from edge effects: marginal regeneration at the bottom or sides of the film and the boundary between the colored and black film at the top. 
As noted above, if the film was rigid instead of mobile (following the distinction from \citet{mysels_soap_1964}), the thickness should follow $h \propto \sqrt{z/t}$ and this drainage scaling would break down. 
For solutions with moderate salinities (up to seawater concentration), this study of film drainage therefore allows us to conclude that (i) the only effect of NaCl on thin film drainage is through viscosity and (ii) relative humidity (and therefore evaporation) does not play a significant role in the drainage, regardless of salinity.

This last remark should be contrasted with the very strong effect that relative humidity has on film lifetime. For a given solution, the lifetime can range from only a few seconds at 30\% humidity to tens of minutes in an almost saturated atmosphere at 90\% humidity. 
In the following we therefore focus on the effect of relative humidity on the lifetime of the films and obtain a scaling law explaining this dependence.

\section{Film lifetime}\label{sec:lifetime}

In this section, we discuss the film lifetime and the effect of convective evaporation on the lifetime statistics and mean values. We rationalize the observations using scaling arguments  as well as measurements of the rupture location in the black film and its evolution.

\subsection{Mean lifetime and convective evaporation}

\begin{figure}
    \centering
    \includegraphics[width=0.45\linewidth]{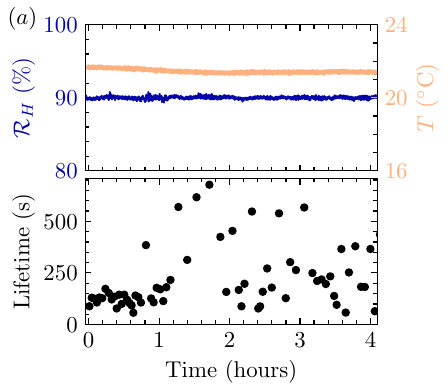}
    \includegraphics[width = 0.452\linewidth]{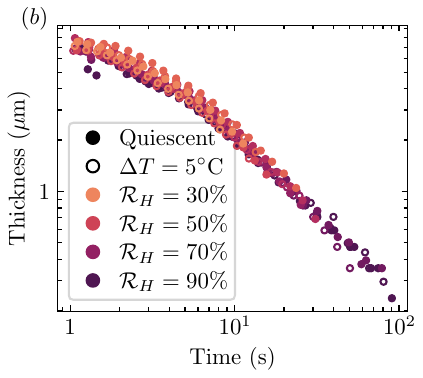}
    \caption{(a) Example of a time series of lifetime measurements without the convection activated in the chamber. After about an hour, the lifetime drastically changes while the parameters (temperature and humidity) remain constant. (top) Relative humidity (blue) and temperature (orange) close to the film. (bottom) Lifetime of the films over the course of the experiment. \\(b) Drainage with a quiescent (full symbols) or mixed atmosphere (empty symbols) at a fixed location in the film. Colors show relative humidity.}
    \label{fig:fluc_lifetime}
\end{figure}

To investigate film lifetime, we repeat the drainage experiment at least a hundred times to obtain robust statistics. In the cases with large lifetime fluctuations (relative humidity $\mathcal{R}_H > 70\%$) the experiment is repeated two hundred times. Given the limited influence of the solution on drainage, we use solutions of Triton X-100 without any salt or glycerol added to focus on the effect of the air around the film through humidity or flow. In particular, we investigate the dependence of lifetime with humidity because of the stark contrast between its lack of effect on drainage and the key role it plays in determining the lifetime of soap films. When the experiment is repeated, a difficulty arises: sometimes the lifetime of the film for a given set of conditions varies drastically throughout repetitions (see Fig. \ref{fig:fluc_lifetime} (a), where the mean lifetime changes by a factor five with no variation of the control parameters). This difficulty, which was previously noted by \citet{poulain_ageing_2018} in the case of surface bubbles is particularly striking at humidities close to saturation. The distribution of lifetimes appears to change drastically and irreversibly towards extremely long times (tens of minutes) even though the experimental conditions do not vary (the temperature and relative humidity are continuously recorded in the chamber, see Fig. \ref{fig:fluc_lifetime} (a)).

These modulations in the distribution of lifetimes are a major source of uncertainty and difficulties in drawing conclusion in studies about surface bubbles. We interpret these fluctuations in our case as being due to slow modulations of the evaporation of the surface from which the film is drawn. As the surface evaporates, the air right above it gets charged in water vapor, locally increasing the humidity. The humidity close to the film is therefore closer to saturation than what would be expected from the far field humidity measurement that controls the humidity in the chamber. Exactly how or why this local humidity increase is modulated remains for now uncertain, one possibility being the formation or destabilization of a plume of humid air: as it is lighter than dry air, a plume of high humidity air forms over the center of the bath \citep{dollet_natural_2017,boulogne_convective_2018}. This plume has not been characterized experimentally, however plumes around evaporating droplets are common \citep{dehaeck_vaporbased_2014,dietrich_role_2016} and the mass transfer problem is in this case completely analogous to the heat transfer problem of a heated disk \citep{lienhardv_heat_2024} where plumes of warm air have been widely studied \citep{torrance_experiments_1969,lopez_instability_2013,kwak_natural_2018,khrapunov_structure_2020}. When we introduce airflow in the chamber, this localized increase in humidity is washed away by the turbulent mixing in the chamber, making the humidity field homogeneous. Using the Rayleigh-Bénard instability (Fig. \ref{fig:setup} (b)), the airflow is weak enough such that it does not disturb the drainage studied in the previous section but strong enough to continuously mix and renew the air in contact with the film. We show a comparison of the drainage curves with and without airflow in Fig. \ref{fig:fluc_lifetime} (b) and demonstrate that the drainage is not modified.

\begin{figure}
    \centering
    \includegraphics[width = \textwidth]{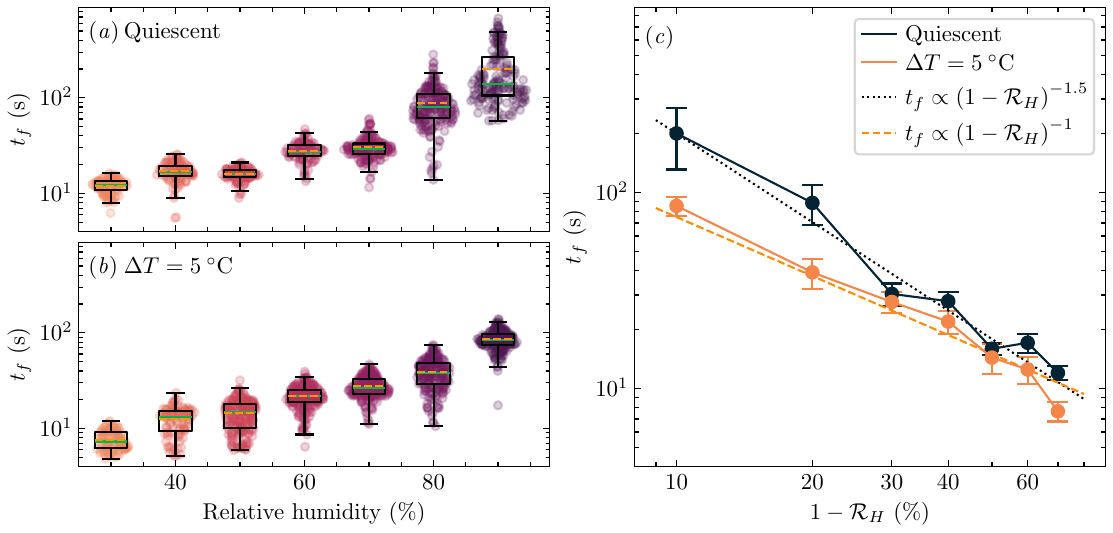}
    \caption{Lifetime distribution and average quantities as a function of relative humidities in the case with (a) a quiescent atmosphere ($\Delta T = 0\:^\circ$C) and (b) with the convection turned on ($\Delta T = 5\:^\circ$C). Each dot is a realization of the experiment (100 - 200 per condition). Boxplots shows averaged quantities: (green line) median, (orange dashed line) mean, (box width) inter-quartile range, (whiskers) furthest data point within 1.5 times inter-quartile range from the box. \quad (c) Evolution of the lifetime with $1 - \mathcal{R}_H$, (black) with a quiescent atmosphere and (orange) with the convection turned on. (Orange dashed line) fit with a power law of exponent -1, (black dotted line) fit with a power law of exponent -1.5.}
    \label{fig:lifetime}
\end{figure}

The resulting data for the lifetime $t_f$ is shown in Fig. \ref{fig:lifetime} as a function of relative humidity, both for a quiescent atmosphere ($\Delta T = 0\:^\circ$C, panel (a)) and with the mixing activated ($\Delta T = 5\:^\circ$C, panel (b)). For each condition, the experiment was repeated at least 100 times to ensure robust statistics (each dot is a single experiment), and for cases with the largest deviations, 200 experiments were performed. In both cases a very clear trend of increasing lifetime with relative humidity can be observed (note the logarithmic vertical scale), with lifetimes ranging from a few seconds at 30\% $\mathcal{R}_H$ up to more than 10 minutes for the longest lasting film at 90\% $\mathcal{R}_H$ in a quiescent atmosphere. 
The effect of forcing the mixing in the chamber is twofold: firstly, it suppresses the anomalous, extremely long-lasting films occurring at high humidity in the quiescent case. As a result the standard deviation $\sigma$ of the lifetime over repeated experiments is reduced: $\sigma(t_f) / \left< t_f \right> \approx 0.3$ throughout the range of humidities when convection is activated but reaches as high as 0.7 without it. Second, mixing shifts the whole distribution towards shorter lifetimes, especially at high humidities. The films are therefore sensitive not only to the humidity in the chamber, but also to how well the atmosphere is mixed within the chamber and the lifetime can be varied by a factor 3 simply by changing the airflow. 

We additionally investigated the dependence of lifetime with the intensity of convection (at 80\% $\mathcal{R}_H$ and up to $\Delta T = 15\:^\circ$C, App. \ref{app:tf_deltat}) but found no significant trend. If no mixing is imposed in the chamber, the lifetime of the object studied is \emph{a priori} sensitive to all the experimental apparatus used: size of bath from which the film is drawn, position and flow rate of the humid and dry air inlets, etc. With any of the temperature differences tested it appears that enough mixing occurs to remove these unwanted dependencies and improve the repeatability of the experiment. As long as this mixing is not strong enough to alter the drainage of the film or directly make it burst, the intensity of the convection appears to play only a weak role in setting the lifetime.

\subsection{Evaporative flux as a function of relative humidity}

We can now synthetize the scaling of the lifetime with the relative humidity. Figure \ref{fig:lifetime} shows the lifetime $t_f$ as a function of $1 - \mathcal{R}_H$ and we can observe that in both the mixed and unmixed cases, the lifetime seems to follow a power law. In the case where the atmosphere is mixed, this power law has a slope close to -1 (dashed line) and in the quiescent case a slope of about -1.5 (dotted line). 
From observations of the film bursting with a high speed camera, the film always seems to burst at its top, in the black film, a common behavior for heavily contaminated bubbles and soap films (\citet{champougny_life_2016,pasquet_lifetime_2024}, see Fig. \ref{fig:exemple_bf} (a)). The black film (visible at the top of Fig. \ref{fig:setup} (b)) is a region too thin to measure with our visible light technique, which may be much thinner than the rest of the film and heavily dependent on the surfactant used \citep{langevin_rupture_2020}. Still, within our experiment, this region of black film always appears only a few seconds after the film has started thinning. Under the hypothesis that its initial thickness $h_\text{BF}$ is always similar and that evaporation is the only mechanism responsible for the removal of fluid from this region, we can obtain a typical lifetime:
\begin{equation}
    t_f \sim \frac{h_\text{BF}}{j_e},
\end{equation}
where $j_e$ is the evaporative flux and can be expressed as $j_e = k_e (c_s - c_\infty) = k_e c_s (1 - \mathcal{R}_H)$ with $k_e$ the mass transfer coefficient, $c_s$ the saturation mass concentration of water in air at a given temperature and $c_\infty = \mathcal{R}_H c_s$ the mass concentration in the far field imposed by the experiment. Finding the evaporative flux therefore requires expressing the mass transfer coefficient as a function of the parameters of the evaporation problem (far field relative humidity, geometry of the setup, background motion in the air of the chamber, etc.).

In the case with a quiescent atmosphere, the mass transfer coefficient $k_e$ needs to be computed taking into account the evaporation of the bath and the whole film, through diffusion and humidity-induced convection. The number measuring the importance of this phenomenon is the Grashof number \citep{dollet_natural_2017}, comparing buoyancy with viscous dissipation :
\begin{equation}
    Gr = \left| \frac{\rho_s - \rho_\infty}{\rho_\infty} \right| \frac{g \mathcal{L}^3}{\nu_a^2},
\end{equation}
with $\rho_s$ and $\rho_\infty$ the air density close to the film (and therefore saturated with humidity) and in the far field, respectively. The length scale $\mathcal{L}$ is in our case of the order of the size of the bath or the film height (both about 1~cm). The Grashof number for this problem is typically $\mathcal{O}(10^4 - 10^5)$. Previous studies have focused on the mass transfer due to evaporation of the bath alone \citep{dollet_natural_2017} and the film alone \citep{boulogne_convective_2018}. In the case of the film alone, it is possible to theoretically find a scaling: $k_e \propto Gr^{1/4}$. 

The Grashof number depends on the relative difference in density between the saturated air and the one in the far field. The air density in incompressible conditions depends on humidity following \citep{tsilingiris_thermophysical_2008}:
\begin{equation}
    \rho(\mathcal{R}_H) = \frac{P_0}{RT}M_d \left( 1 - \mathcal{R}_H \left( 1 - \frac{M_w}{M_d} \right) \frac{P_s}{P_0}\right),
\end{equation}
with $P_0$ the atmospheric pressure, $P_s$ the saturated partial pressure of water vapor, $R$ the ideal gas constant, $T$ the temperature, and $M_w$ and $M_d$ the molar masses of water vapor and dry air, respectively. The relative difference of density therefore reads:
\begin{equation}
    \frac{\rho_s - \rho_\infty}{\rho_\infty} = \frac{\rho(\mathcal{R}_H = 0)}{\rho_\infty} \left( 1 - \frac{M_w}{M_d} \right) \frac{P_s}{P_0} \left( 1 - \mathcal{R}_H \right).
\end{equation}
The relative difference of density is very small (less than 1\% between dry and saturated air), the Grashof number can therefore be written as:
\begin{equation}
    Gr(\mathcal{R}_H) = (1 - \mathcal{R}_H) \times \underbrace{\left( 1 - \frac{M_w}{M_d} \right) \frac{P_s}{P_0} \frac{g \mathcal{L}^3}{\nu_a^2}}_{Gr_0}.
\end{equation}

Finally, the dependence of the mass transfer coefficient with relative humidity can be put forward: $k_e \propto (1 - \mathcal{R}_H)^{1/4}$ in the case of a film without a bath, giving $t_f \propto (1 - \mathcal{R}_H)^{-5/4}$. 

In the case of our problem, because of the geometry combining the evaporation of the bath and the film, we do not know the relation between $k_e$ and $Gr$. We make the hypothesis that a similar scaling to the one we found for a film without a bath exists, but with an exponent $\alpha$ that may be different (i.e. $k_e \propto Gr^\alpha$). We therefore expect $t_f \propto (1 - \mathcal{R}_H)^{-(1 + \alpha)}$ and find empirically that $\alpha = 0.5$ is a good fit of our data (black dotted line of Fig. \ref{fig:lifetime}). This scaling is the result of the non-trivial flow generated by the combined evaporation of the bath and the film at the same time.

In the case where mixing is activated, finding the mass transfer coefficient is simpler as the convection is forced. In that case, we make the hypothesis that the mass transfer coefficient $k_e$ is independent of the relative humidity and controlled by the turbulent flow, only depending on the velocity resulting from the Rayleigh-Bénard convection $u_\textsc{rb}$ (i.e. the evaporation flux associated with natural convection is assumed to be negligible compared to the one associated with forced convection), {\it i.e.} $k_e = f(Re, Sc$), with $Re = \rho_a u_\textsc{rb} W / \mu_a$ the Reynolds number associated with the flow over the film, and $Sc = \nu_a / \mathcal{D}$ the Schmidt number comparing kinematic viscosity with the mass diffusivity of water vapor $\mathcal{D}$ \citep{lienhardv_heat_2024}. As a result, we obtain the scaling
\begin{equation}
    t_f \propto (1 - \mathcal{R}_H)^{-1},
\end{equation}
which matches well with the experiments (dashed orange line in Fig. \ref{fig:lifetime}). Additionally, the effect of changing the mixing intensity is reflected through the Reynolds number dependence of the mass transfer coefficient. From the classical analogy between heat and mass transfer \citep{lienhardv_heat_2024}, the mass transfer coefficient scales for flows over a flat surface varies with the Reynolds number with a power smaller than unity. The range of velocities accessible with our Rayleigh-Bénard setup (3 mm/s for $\Delta T = 3^\circ$C and 6 mm/s for $\Delta T = 10^\circ$C, measured using particle image velocimetry, see App. \ref{app:tf_deltat}) is therefore most likely too small to be noticeable with the still important lifetime fluctuations of the films.

We have therefore demonstrated that the film lifetime is better defined when the air around the film is mixed. As the evaporation rate controls the bursting time, the lifetime of the film is very sensitive to local fluctuations of humidity. Adding a background air flow ensures that the humidity is the same everywhere, improving the statistics of film lifetime, and reduces the overall film lifetime.

\subsection{Evolution of the black film extent}

\begin{figure}
    \centering
    \includegraphics[width = \textwidth]{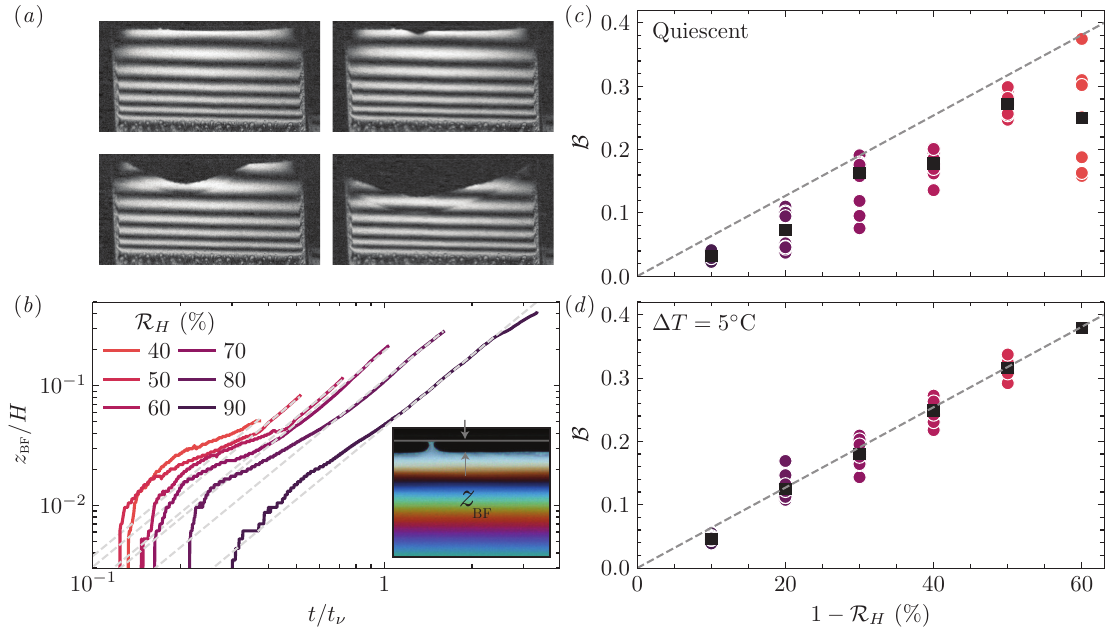}
    \caption{(a) Chronophotography of a film bursting. The film is illuminated by a laser, the frames are spaced by 150~\textmu s. \quad (b) Vertical extent of the black film $z_\textsc{BF}$ over time. Colors indicate relative humidity. Dashed gray lines are fits with a power law (Eq. \eqref{eq:zbf}) fitted onto the region where $z_\textsc{BF} > 1$~mm. The data presented correspond to a case with $\Delta T = 5^\circ C$. \quad (c) Prefactor $\mathcal{B}$ (Eq. \ref{eq:zbf}) as a function of $1 - \mathcal{R}_H$, in a quiescent atmosphere and with convection turned on. Colored markers are individual experiments and black markers are averages over repetitions. (dashed lines) Linear fit $\mathcal{B} = 0.635\times(1 - \mathcal{R}_H)$.}
    \label{fig:exemple_bf}
\end{figure}

From high speed observations of the film bursting (Fig. \ref{fig:exemple_bf} (a)), we can notice that the burst location is always at the top: where the film is the thinnest and where the black film is located \citep{saulnier_study_2014,pasquet_lifetime_2024}. To link our observations on lifetime and drainage, we therefore investigate the time evolution of this region of the film, which usually is small enough to not alter the drainage but appears to play an important role for the lifetime. In particular, we aim to put forward its dependence on humidity and the mixing of the atmosphere in the chamber. Because the thickness of this region is not accessible with our visible light visualization method, we can only probe its spatial extent within the frame. As the film ages, the black film grows and takes up more and more space. For extremely long-lasting films (with a quiescent atmosphere), it may fill almost half the frame. In Fig. \ref{fig:exemple_bf} (b), we show examples of the evolution of the vertical extent of the black film $z_\textsc{bf}$ for each relative humidity condition and with convection in the atmosphere. We measure the vertical extent of the film in the middle where it is the largest. The black film grows continuously throughout the lifetime of the films at a rate that increases with time. In contrast to the drainage, the extent of the black film depends on relative humidity: it grows much faster in a dry atmosphere, and does so from the very beginning of the lifetime of the films. 
At long times, we find empirically that the evolution of the black film follows closely a power law with an exponent of~2: 
\begin{equation}\label{eq:zbf}
    z_\textsc{bf} / H = \mathcal{B}\times (t / t_\nu)^2,
\end{equation}
with $\mathcal{B}$ a dimensionless constant that \emph{a priori} depends on humidity, convection and the solution used (gray dashed lines). This fit describes well the evolution of $z_\textsc{bf}$ after a transition period that typically lasts until $z_\textsc{bf} > 1$~mm (or $z_\textsc{bf} / H > 0.05$). At low humidities, the films do not last long enough for this regime to be visible. At high humidities the growth of the black film slows down for very long times (top right, when $t / t_0 \gtrsim 10$) when its extent reaches a significant portion of the total frame.


The dependence of the prefactor $\mathcal{B}$ with relative humidity is shown in Fig. \ref{fig:exemple_bf} (c), both for a quiescent atmosphere and one where the convection is on. In both cases, $\mathcal{B}$ grows with $1 - \mathcal{R}_H$, \emph{i.e.} the black film grows faster when the humidity is low. Comparing the cases where the convection is activated or not, we can notice that the dispersion of the data is much smaller when the air is mixed and in that case $\mathcal{B}$ is simply proportional to $1 - \mathcal{R}_H$ (bottom panel). This result indicates that the growth rate of the black film is a good indicator of the local humidity conditions close to the top of the film. 

When there is no convection (top panel) the dispersion is important but the rate at which the film grows is systematically smaller than in the well mixed case. The local humidity conditions close to the black film are different than the ones set in the far field by our control system: for instance, the markers at 60\% humidity ($1 - \mathcal{R}_H = 40$~\%) most likely were recorded with a humidity close to the film on average equal to 70~\%. The evaporation of the bath increases the humidity close to the film, slowing down the growth of the black film and extending the lifetime of the whole film. 

\section{Conclusion}

We studied the effect of salt and humidity on the drainage and lifetime of mobile soap films. By measuring the thickness of the films over time, we were able to show that (i) the drainage of the film is independent of the humidity of the atmosphere, (ii) up to twice the salt concentration of seawater, the only effect of NaCl on the drainage of thin films is through viscosity, and (iii) the thickness of the film takes the form $h(z, t) \propto z / t$ for all the conditions tested, from a few seconds after the start of the experiment and outside the marginal regeneration zone. This last formula lacks a theoretical derivation for now; a major source of uncertainty is the role played by the marginal regeneration at the bottom and along the side edges of the film.

Relative humidity dictates the life of the film: the local humidity near the film alters the evaporation flux and the evolution of the black film, which leads to the bursting of the film. By mixing the air, the long-term variability of the local conditions is eliminated, and the humidity field becomes homogeneous. As a result, the fluctuations of the lifetime and its mean value are reduced, this is especially true at high humidities where large, long term lifetime fluctuations are present if there is no imposed flow. If left uncontrolled, the local conditions are sensitive to the experimental setup (size of the bath, positioning of the air inlet, etc.) instead of being dictated by the far-field imposed humidity value. In a very still atmosphere, it is possible to produce an extremely long-lived film that is instantly burst by the slightest breeze. It is still unclear what about the local increase of the humidity field causes important fluctuations in lifetime, especially those visible in Fig. \ref{fig:fluc_lifetime} (a). A possible culprit might be a transition in the convective regime near the film, visualizations of the humidity field similar to the ones in \cite{dehaeck_vaporbased_2014} or \cite{dietrich_role_2016} might shed light on this puzzling phenomenon in the future. 

A very natural perspective of this work is the extension to the case of surface bubbles. We are currently exploring whether the presence of airflow over the surface influences the drainage and / or lifetime of bubbles.

\textbf{Acknowledgement}
This work was supported by NSF grant 2242512, NSF CAREER 1844932 and the Cooperative Institute for Modeling the Earth's System at Princeton University to L.D.

\textbf{Declaration of Interests}
The authors report no conflict of interest.

\appendix

\section{Langmuir through measurements}\label{app:langmuir}

Every solution used in this study was tested in a Langmuir trough (KSV NIMA, model KN 1003) using the Wilhelmy plate method to record its surface tension isotherm. We record the dynamic properties of the surface by setting aside a small amount of solution and placing it in the trough, letting it rest until the surface tension does not evolve significantly. The surface tension is measured with a platinum Wilhelmy plate. After that, two Teflon barriers compress the surface at a rate of 270~mm/min allowing us to obtain the surface tension $\gamma$ as a function of the through area. 

All the curves used for the soap films ([Triton X-100] = 3 CMC and various salinities in shades of blue to yellow) show typical behavior for a solution with a surfactant at a concentration above the CMC: a constant low value (here equal to 32~mN/m). NaCl concentration has only a very minor effect on surface tension: the effect in impossible to discern for the static surface tension and at high compression, higher salt concentration slightly reduce surface tension.
\begin{figure}
    \centering
    \includegraphics[width = 0.6\linewidth]{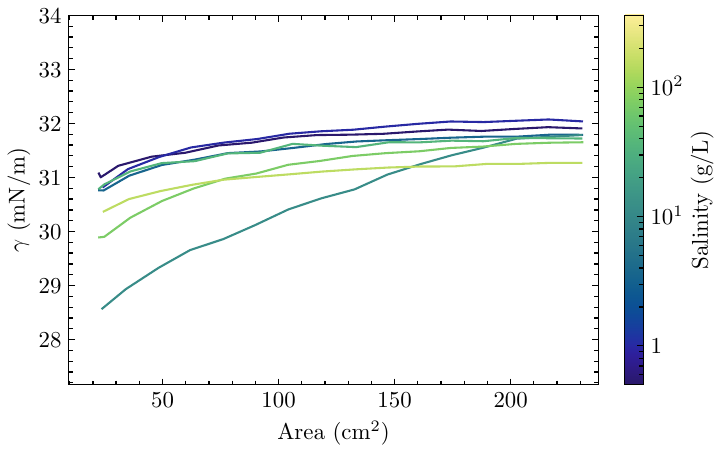}
    \caption{Surface tension as a function of through area for Triton X-100 at 3 CMC with different NaCl concentrations.}
    \label{fig:langmuir}
\end{figure}

\section{Lifetime dependence on mixing intensity}\label{app:tf_deltat}

\begin{figure}
    \centering
    \includegraphics[width = 0.6\linewidth]{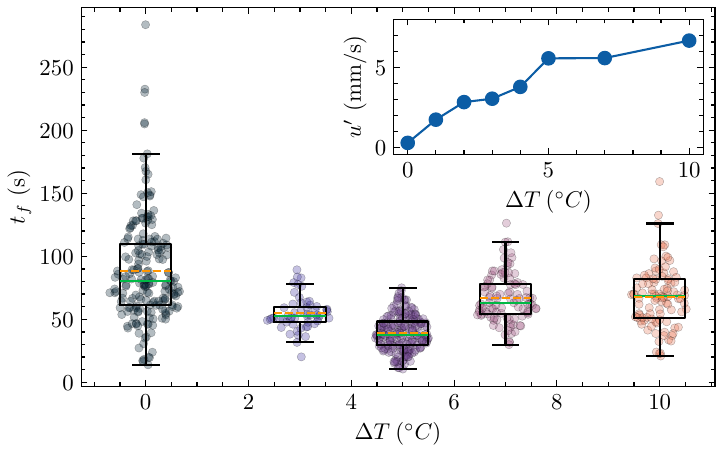}
    \caption{Film lifetime as a function of the imposed $\Delta T$ with a relative humidity of 80\%. (Inset) root-mean-square velocity of the air over the surface as a function of $\Delta T$.}
    \label{fig:lifetime_deltat}
\end{figure}

We additionally investigated the dependence of lifetime with the intensity of the convection in the chamber, measured through the temperature difference between the top and bottom $\Delta T$. The results are shown in Fig. \ref{fig:lifetime_deltat} for a humidity of 80\%. First, the anomalously long lifetimes of several hundred seconds are only visible in the quiescent case. The smallest temperature difference (3~$^\circ C$), corresponding to a velocity of about 4~mm/s is enough to suppress the large lifetime fluctuations and the standard deviation of the lifetime is greatly reduced.
Finally, there is no clear dependence of mean lifetime with the intensity of the convection. This might mean that the atmosphere is sufficiently mixed from the smallest temperature difference and that the influence of the flow rate over the film is weak in this range of velocities.


\begin{thebibliography}{}

    \bibitem[Anguelova and Huq, 2017]{anguelova_effects_2017}
    Anguelova, M.~D. and Huq, P. (2017).
    \newblock Effects of {{Salinity}} on {{Surface Lifetime}} of {{Large Individual
      Bubbles}}.
    \newblock {\em Journal of Marine Science and Engineering}, 5(3):41.
    
    \bibitem[Atkins and Elliott, 2010]{atkins_investigating_2010}
    Atkins, L.~J. and Elliott, R.~C. (2010).
    \newblock Investigating thin film interference with a digital camera.
    \newblock {\em American Journal of Physics}, 78(12):1248--1253.
    
    \bibitem[Berg et~al., 2005]{berg_experimental_2005}
    Berg, S., Adelizzi, E.~A., and Troian, S.~M. (2005).
    \newblock Experimental {{Study}} of {{Entrainment}} and {{Drainage Flows}} in
      {{Microscale Soap Films}}.
    \newblock {\em Langmuir}, 21(9):3867--3876.
    
    \bibitem[Berny et~al., 2021]{berny_statistics_2021}
    Berny, A., Popinet, S., S{\'e}on, T., and Deike, L. (2021).
    \newblock Statistics of {{Jet Drop Production}}.
    \newblock {\em Geophysical Research Letters}, 48(10):e2021GL092919.
    
    \bibitem[Boulogne, 2019]{boulogne_cheap_2019}
    Boulogne, F. (2019).
    \newblock Cheap and versatile humidity regulator for environmentally controlled
      experiments.
    \newblock {\em The European Physical Journal E}, 42(4):51.
    
    \bibitem[Boulogne and Dollet, 2018]{boulogne_convective_2018}
    Boulogne, F. and Dollet, B. (2018).
    \newblock Convective evaporation of vertical films.
    \newblock {\em Soft Matter}, 14(9):1665--1671.
    
    \bibitem[Burgess et~al., 1999]{burgess_instability_1999}
    Burgess, J.~M., Bizon, C., McCormick, W.~D., Swift, J.~B., and Swinney, H.~L.
      (1999).
    \newblock Instability of the {{Kolmogorov}} flow in a soap film.
    \newblock {\em Physical Review E}, 60(1):715--721.
    
    \bibitem[Champougny et~al., 2018]{champougny_influence_2018}
    Champougny, L., Miguet, J., Henaff, R., Restagno, F., Boulogne, F., and Rio, E.
      (2018).
    \newblock Influence of {{Evaporation}} on {{Soap Film Rupture}}.
    \newblock {\em Langmuir}, 34(10):3221--3227.
    
    \bibitem[Champougny et~al., 2016]{champougny_life_2016}
    Champougny, L., Roch{\'e}, M., Drenckhan, W., and Rio, E. (2016).
    \newblock Life and death of not so ``bare'' bubbles.
    \newblock {\em Soft Matter}, 12(24):5276--5284.
    
    \bibitem[Champougny et~al., 2015]{champougny_surfactantinduced_2015}
    Champougny, L., Scheid, B., Restagno, F., Vermant, J., and Rio, E. (2015).
    \newblock Surfactant-induced rigidity of interfaces: A unified approach to free
      and dip-coated films.
    \newblock {\em Soft Matter}, 11(14):2758--2770.
    
    \bibitem[Cunliffe et~al., 2013]{cunliffe_sea_2013}
    Cunliffe, M., Engel, A., Frka, S., Ga{\v s}parovi{\'c}, B., Guitart, C.,
      Murrell, J.~C., Salter, M., Stolle, C., {Upstill-Goddard}, R., and Wurl, O.
      (2013).
    \newblock Sea surface microlayers: {{A}} unified physicochemical and biological
      perspective of the air--ocean interface.
    \newblock {\em Progress in Oceanography}, 109:104--116.
    
    \bibitem[{de Gennes}, 2001]{degennes_young_2001}
    {de Gennes}, P.~G. (2001).
    \newblock ``{{Young}}'' {{Soap Films}}.
    \newblock {\em Langmuir}, 17(8):2416--2419.
    
    \bibitem[Dehaeck et~al., 2014]{dehaeck_vaporbased_2014}
    Dehaeck, S., Rednikov, A., and Colinet, P. (2014).
    \newblock Vapor-{{Based Interferometric Measurement}} of {{Local Evaporation
      Rate}} and {{Interfacial Temperature}} of {{Evaporating Droplets}}.
    \newblock {\em Langmuir}, 30(8):2002--2008.
    
    \bibitem[Deike, 2022]{deike_mass_2022}
    Deike, L. (2022).
    \newblock Mass {{Transfer}} at the {{Ocean}}--{{Atmosphere Interface}}: {{The
      Role}} of {{Wave Breaking}}, {{Droplets}}, and {{Bubbles}}.
    \newblock {\em Annual Review of Fluid Mechanics}, 54(1):191--224.
    
    \bibitem[Deike et~al., 2022]{deike_mechanistic_2022}
    Deike, L., Reichl, B.~G., and Paulot, F. (2022).
    \newblock A {{Mechanistic Sea Spray Generation Function Based}} on the {{Sea
      State}} and the {{Physics}} of {{Bubble Bursting}}.
    \newblock {\em AGU Advances}, 3(6):e2022AV000750.
    
    \bibitem[Dietrich et~al., 2016]{dietrich_role_2016}
    Dietrich, E., Wildeman, S., Visser, C.~W., Hofhuis, K., Kooij, E.~S.,
      Zandvliet, H. J.~W., and Lohse, D. (2016).
    \newblock Role of natural convection in the dissolution of sessile droplets.
    \newblock {\em Journal of Fluid Mechanics}, 794:45--67.
    
    \bibitem[Dollet and Boulogne, 2017]{dollet_natural_2017}
    Dollet, B. and Boulogne, F. (2017).
    \newblock Natural convection above circular disks of evaporating liquids.
    \newblock {\em Physical Review Fluids}, 2(5):053501.
    
    \bibitem[Drazin, 2002]{drazin_rayleighbenard_2002}
    Drazin, P.~G. (2002).
    \newblock Rayleigh-{{B{\'e}nard Convection}}.
    \newblock In {\em Introduction to {{Hydrodynamic Stability}}}, Cambridge
      {{Texts}} in {{Applied Mathematics}}, pages 93--122. Cambridge University
      Press, Cambridge.
    
    \bibitem[Dubitsky et~al., 2023]{dubitsky_effects_2023}
    Dubitsky, L., Stokes, M.~D., Deane, G.~B., and Bird, J.~C. (2023).
    \newblock Effects of {{Salinity Beyond Coalescence}} on {{Submicron Aerosol
      Distributions}}.
    \newblock {\em Journal of Geophysical Research: Atmospheres},
      128(10):e2022JD038222.
    
    \bibitem[{El-Dessouky} et~al., 2002]{el-dessouky_evaporation_2002}
    {El-Dessouky}, H.~T., Ettouney, H.~M., Alatiqi, I.~M., and {Al-Shamari}, M.~A.
      (2002).
    \newblock Evaporation {{Rates}} from {{Fresh}} and {{Saline Water}} in {{Moving
      Air}}.
    \newblock {\em Industrial \& Engineering Chemistry Research}, 41(3):642--650.
    
    \bibitem[Jiang et~al., 2022]{jiang_submicron_2022}
    Jiang, X., Rotily, L., Villermaux, E., and Wang, X. (2022).
    \newblock Submicron drops from flapping bursting bubbles.
    \newblock {\em Proceedings of the National Academy of Sciences},
      119(1):e2112924119.
    
    \bibitem[Jiang et~al., 2024]{jiang_abyss_2024}
    Jiang, X., Rotily, L., Villermaux, E., and Wang, X. (2024).
    \newblock Abyss {{Aerosols}}: {{Drop Production}} from {{Underwater Bubble
      Collisions}}.
    \newblock {\em Physical Review Letters}, 133(2):024001.
    
    \bibitem[Kestin et~al., 1981]{kestin_tables_1981}
    Kestin, J., Khalifa, H.~E., and Correia, R.~J. (1981).
    \newblock Tables of the dynamic and kinematic viscosity of aqueous {{NaCl}}
      solutions in the temperature range 20--150 {$^\circ$}{{C}} and the pressure
      range 0.1--35 {{MPa}}.
    \newblock {\em Journal of Physical and Chemical Reference Data}, 10(1):71--88.
    
    \bibitem[Kharlamova et~al., 2024]{kharlamova_interfacetemplated_2024}
    Kharlamova, A., Boulogne, F., Fontaine, P., Rouzi{\`e}re, S., Hemmerle, A.,
      Goldmann, M., and Salonen, A. (2024).
    \newblock Interface-{{Templated Crystal Growth}} in {{Sodium Dodecyl Sulfate
      Solutions}} with {{NaCl}}.
    \newblock {\em Langmuir}, 40(1):84--90.
    
    \bibitem[Khrapunov and Chumakov, 2020]{khrapunov_structure_2020}
    Khrapunov, E. and Chumakov, Y. (2020).
    \newblock Structure of the natural convective flow above to the horizontal
      surface with localized heating.
    \newblock {\em International Journal of Heat and Mass Transfer}, 152:119492.
    
    \bibitem[Kwak et~al., 2018]{kwak_natural_2018}
    Kwak, D.-B., Noh, J.-H., and Yook, S.-J. (2018).
    \newblock Natural convection flow around heated disk in cubical enclosure.
    \newblock {\em Journal of Mechanical Science and Technology}, 32(5):2377--2384.
    
    \bibitem[Langevin, 2020]{langevin_rupture_2020}
    Langevin, D. (2020).
    \newblock On the rupture of thin films made from aqueous surfactant solutions.
    \newblock {\em Advances in Colloid and Interface Science}, 275:102075.
    
    \bibitem[Langevin and Sonin, 1994]{langevin_thinning_1994}
    Langevin, D. and Sonin, A.~A. (1994).
    \newblock Thinning of soap films.
    \newblock {\em Advances in Colloid and Interface Science}, 51:1--27.
    
    \bibitem[Lewis and Schwartz, 2004]{lewis_sea_2004}
    Lewis, R. and Schwartz, E. (2004).
    \newblock {\em Sea {{Salt Aerosol Production}}: {{Mechanisms}}, {{Methods}},
      {{Measurements}} and {{Models}}---{{A Critical Review}}}, volume 152 of {\em
      Geophysical {{Monograph Series}}}.
    \newblock American Geophysical Union, Washington, D. C.
    
    \bibitem[Lhuissier and Villermaux, 2012]{lhuissier_bursting_2012}
    Lhuissier, H. and Villermaux, E. (2012).
    \newblock Bursting bubble aerosols.
    \newblock {\em Journal of Fluid Mechanics}, 696:5--44.
    
    \bibitem[Lienhard and Lienhard, 2024]{lienhardv_heat_2024}
    Lienhard, V, J.~H. and Lienhard, IV, J.~H. (2024).
    \newblock {\em A {{Heat Transfer Textbook}}}.
    \newblock Phlogiston Press, Cambridge, MA, 6th edition.
    
    \bibitem[Lopez and Marques, 2013]{lopez_instability_2013}
    Lopez, J.~M. and Marques, F. (2013).
    \newblock Instability of plumes driven by localized heating.
    \newblock {\em Journal of Fluid Mechanics}, 736:616--640.
    
    \bibitem[M{\aa}rtensson et~al., 2003]{martensson_laboratory_2003}
    M{\aa}rtensson, E.~M., Nilsson, E.~D., {de Leeuw}, G., Cohen, L.~H., and
      Hansson, H.-C. (2003).
    \newblock Laboratory simulations and parameterization of the primary marine
      aerosol production.
    \newblock {\em Journal of Geophysical Research: Atmospheres}, 108(D9).
    
    \bibitem[May et~al., 2016]{may_lake_2016}
    May, N.~W., Axson, J.~L., Watson, A., Pratt, K.~A., and Ault, A.~P. (2016).
    \newblock Lake spray aerosol generation: A method for producing representative
      particles from freshwater wave breaking.
    \newblock {\em Atmospheric Measurement Techniques}, 9(9):4311--4325.
    
    \bibitem[Mazzatenta et~al., 2024]{mazzatenta_linking_2024}
    Mazzatenta, M., Erinin, M.~A., N{\'e}el, B., and Deike, L. (2024).
    \newblock Linking emitted drops to collective bursting bubbles across a wide
      range of bubble size distributions.
    
    \bibitem[Monier et~al., 2024]{monier_selfsimilar_2024}
    Monier, A., Gauci, F.-X., Claudet, C., Celestini, F., Brouzet, C., and
      Raufaste, C. (2024).
    \newblock Self-similar and universal dynamics in drainage of mobile soap films.
    \newblock {\em Physical Review Fluids}, 9(12):124001.
    
    \bibitem[Mysels, 1959]{mysels_soap_1959}
    Mysels, K.~J. (1959).
    \newblock {\em Soap {{Films}}: {{Studies}} of {{Their Thinning}} and a
      {{Bibliography}}}.
    \newblock Pergamon Press.
    
    \bibitem[Mysels, 1964]{mysels_soap_1964}
    Mysels, K.~J. (1964).
    \newblock Soap {{Films}} and {{Some Problems}} in {{Surface}} and {{Colloid
      Chemistry}} {\textsuperscript{1}}.
    \newblock {\em The Journal of Physical Chemistry}, 68(12):3441--3448.
    
    \bibitem[Naire et~al., 2000]{naire_insoluble_2000}
    Naire, S., Braun, R., and Snow, S. (2000).
    \newblock An {{Insoluble Surfactant Model}} for a {{Vertical Draining Free
      Film}}.
    \newblock {\em Journal of Colloid and Interface Science}, 230(1):91--106.
    
    \bibitem[Park et~al., 2014]{park_mixing_2014}
    Park, J.~Y., Lim, S., and Park, K. (2014).
    \newblock Mixing {{State}} of {{Submicrometer Sea Spray Particles Enriched}} by
      {{Insoluble Species}} in {{Bubble-Bursting Experiments}}.
    \newblock {\em Journal of Atmospheric and Oceanic Technology}, 31(1):93--104.
    
    \bibitem[Pasquet et~al., 2024]{pasquet_lifetime_2024}
    Pasquet, M., Boulogne, F., Restagno, F., and Rio, E. (2024).
    \newblock Lifetime of vertical giant soap films: Role of the relative humidity
      and film dimensions.
    \newblock {\em Soft Matter}, page 10.1039.D3SM01629C.
    
    \bibitem[Pasquet et~al., 2022]{pasquet_impact_2022}
    Pasquet, M., Boulogne, F., {Sant-Anna}, J., Restagno, F., and Rio, E. (2022).
    \newblock The impact of physical-chemistry on film thinning in surface bubbles.
    \newblock {\em Soft Matter}, 18(24):4536--4542.
    
    \bibitem[Poulain et~al., 2018]{poulain_ageing_2018}
    Poulain, S., Villermaux, E., and Bourouiba, L. (2018).
    \newblock Ageing and burst of surface bubbles.
    \newblock {\em Journal of Fluid Mechanics}, 851:636--671.
    
    \bibitem[Roux et~al., 2022]{roux_everlasting_2022a}
    Roux, A., Duchesne, A., and Baudoin, M. (2022).
    \newblock Everlasting bubbles and liquid films resisting drainage, evaporation,
      and nuclei-induced bursting.
    \newblock {\em Physical Review Fluids}, 7(1):L011601.
    
    \bibitem[Saulnier et~al., 2014]{saulnier_study_2014}
    Saulnier, L., Champougny, L., Bastien, G., Restagno, F., Langevin, D., and Rio,
      E. (2014).
    \newblock A study of generation and rupture of soap films.
    \newblock {\em Soft Matter}, 10(16):2899.
    
    \bibitem[Seimiya and Seimiya, 2021]{seimiya_revisiting_2021}
    Seimiya, T. and Seimiya, T. (2021).
    \newblock Revisiting the ``pearl string'' in draining soap bubble film first
      witnessed by {{Sir James Dewar}} some 100 years ago: {{A}} note of analyses
      for the phenomena with related findings.
    \newblock {\em Physics of Fluids}, 33(10):104102.
    
    \bibitem[Seiwert et~al., 2017]{seiwert_velocity_2017}
    Seiwert, J., Kervil, R., Nou, S., and Cantat, I. (2017).
    \newblock Velocity {{Field}} in a {{Vertical Foam Film}}.
    \newblock {\em Physical Review Letters}, 118(4):048001.
    
    \bibitem[Sett et~al., 2013]{sett_gravitational_2013}
    Sett, S., {Sinha-Ray}, S., and Yarin, A.~L. (2013).
    \newblock Gravitational {{Drainage}} of {{Foam Films}}.
    \newblock {\em Langmuir}, 29(16):4934--4947.
    
    \bibitem[Shabalina et~al., 2019]{shabalina_rayleightaylorlike_2019}
    Shabalina, E., B{\'e}rut, A., Cavelier, M., {Saint-Jalmes}, A., and Cantat, I.
      (2019).
    \newblock Rayleigh-{{Taylor-like}} instability in a foam film.
    \newblock {\em Physical Review Fluids}, 4(12):124001.
    
    \bibitem[Shaw and Deike, 2024]{shaw_film_2024}
    Shaw, D.~B. and Deike, L. (2024).
    \newblock Film drop production over a wide range of liquid conditions.
    \newblock {\em Physical Review Fluids}, 9(3):033602.
    
    \bibitem[Torrance et~al., 1969]{torrance_experiments_1969}
    Torrance, K.~E., Orloff, L., and Rockett, J.~A. (1969).
    \newblock Experiments on natural convection in enclosures with localized
      heating from below.
    \newblock {\em Journal of Fluid Mechanics}, 36(1):21--31.
    
    \bibitem[Tsilingiris, 2008]{tsilingiris_thermophysical_2008}
    Tsilingiris, P. (2008).
    \newblock Thermophysical and transport properties of humid air at temperature
      range between 0 and 100{$^\circ$}{{C}}.
    \newblock {\em Energy Conversion and Management}, 49(5):1098--1110.
    
    \bibitem[Veron, 2015]{veron_ocean_2015}
    Veron, F. (2015).
    \newblock Ocean {{Spray}}.
    \newblock {\em Annual Review of Fluid Mechanics}, 47(Volume 47, 2015):507--538.
    
    \bibitem[Zinke et~al., 2022]{zinke_effect_2022}
    Zinke, J., Nilsson, E.~D., Zieger, P., and Salter, M.~E. (2022).
    \newblock The {{Effect}} of {{Seawater Salinity}} and {{Seawater Temperature}}
      on {{Sea Salt Aerosol Production}}.
    \newblock {\em Journal of Geophysical Research: Atmospheres},
      127(16):e2021JD036005.
    
    \end{thebibliography}
\end{document}